\documentclass[twocolumn]{aastex631}
\newcommand{\msun}{M$_{\odot}$}

\definecolor{peach}{rgb}{1.0, 0.498, 0.314}

\definecolor{somecolor}{rgb}{0.8, 0.5, 1.0}

\begin{document}

\title{Young Star Clusters Dominate the Production of Detached Black Hole-Star Binaries}

\author[0000-0003-2654-5239]{Ugo Niccolò Di Carlo}
 \email{udicarlo@sissa.it}
\affiliation{
    Scuola Internazionale Superiore di Studi Avanzati (SISSA)
    Via Bonomea 265, I-34136 Trieste, Italy
}
\affiliation{
    Department of Physics and Astronomy,
    University of North Carolina at Chapel Hill,
    120 E.~Cameron Ave, Chapel Hill, NC, 27599, USA
}
\affiliation{McWilliams Center for Cosmology,
    Department of Physics,
    Carnegie Mellon University,
    5000 Forbes Avenue, Pittsburgh, PA 15213
}
\author[0000-0002-1135-984X]{Poojan Agrawal}
\affiliation{
    Department of Physics and Astronomy,
    University of North Carolina at Chapel Hill,
    120 E.~Cameron Ave, Chapel Hill, NC, 27599, USA
}
\affiliation{McWilliams Center for Cosmology,
    Department of Physics,
    Carnegie Mellon University,
    5000 Forbes Avenue, Pittsburgh, PA 15213
}

\author[0000-0003-4175-8881]{Carl L. Rodriguez}
\affiliation{
    Department of Physics and Astronomy,
    University of North Carolina at Chapel Hill,
    120 E.~Cameron Ave, Chapel Hill, NC, 27599, USA
}
\affiliation{McWilliams Center for Cosmology,
    Department of Physics,
    Carnegie Mellon University,
    5000 Forbes Avenue, Pittsburgh, PA 15213
}

\author[0000-0001-5228-6598]{Katelyn Breivik}
\affiliation{Center for Computational Astrophysics, Flatiron Institute, 162 Fifth Ave, New York, NY, 10010, USA}

\begin{abstract}

The recent discovery of two detached black hole-star (BH-star) binaries from Gaia's third data release has sparkled interest in understanding the formation mechanisms of these systems.  
We investigate the formation of these systems by dynamical processes in young open star clusters (SCs) and via isolated binary (IB) evolution, using a combination of direct $N$-body models and population synthesis simulations.  By comparing dynamical and isolated systems created using the same model of binary stellar evolution, we find that dynamical formation in SCs is nearly 40 times as efficient per unit of star formation at producing BH-star binaries compared to IB evolution. We expand this analysis to the full Milky Way (MW) using a \textit{FIRE-2} hydrodynamical simulation of a MW-mass galaxy.  Even assuming that only $10\%$ of star formation goes into SCs, we find that approximately 4 out of every 5 BH-star systems are formed dynamically, and that the MW contains a total of $\sim 2 \times 10^5$ BH-star systems. Many of these dynamically-formed systems have larger orbital periods, eccentricities, and black hole masses than their isolated counterparts.  For binaries older than 100 Myr, we show that any detectable system with $e\gtrsim0.5$ or $M_{\rm BH}\gtrsim 10\,\mathrm{M_{\odot}}$ can \emph{only} be formed through dynamical processes.  Our MW model predicts between 61 and 210 such detections from the complete DR4 Gaia catalog, with the majority of systems being dynamically formed in massive and metal-rich SCs. Finally, we compare our populations to the recently discovered Gaia BH1 and Gaia BH2, and conclude that the dynamical scenario is the most favorable formation pathway for both systems.

\end{abstract}


\section{Introduction} \label{sec:intro}

Current estimates anticipate that the Milky Way (MW) harbors a population of $\sim10^7-10^9$ stellar black holes (BHs) and $\sim10^4$ massive stars which are likely BH progenitors \citep{garmany1982, reed2003}. While the exact binary fraction of stellar BHs is unknown, the high binary fraction of BH-progenitor stars \citep[e.g.,][]{sana2012} suggests that a substantial quantity of their BH descendants likely exist in binaries as well.
Most of the current observational evidence for BH binary systems in the MW comes from X-ray binary systems, consisting of 20 dynamically confirmed BHs in X-ray binaries, with an additional $\sim50$ X-ray sources identified as strong candidates for containing BHs \citep{mcclintock2006, remillard2006, corralsantana2016}. However, population synthesis models indicate that the majority of BHs in binary systems within the MW are likely to be dormant BHs, with large orbital periods that preclude their involvement in X-ray binaries \citep{portegieszwart1997, corralsantana2016, breivik2017, chawla2022}, making them harder to be detected.
Over the past few years, it has become possible to identify dormant BHs through the motion of their luminous companions, with the first of these binaries being identified through the radial velocity of their luminous components in the globular cluster (GC) NGC 3201 \citep{giesers2018, giesers2019}.  More recently, proper motion data from ESO Gaia's Data Release 3 \citep[DR3,][]{gaia2022c} has enabled the discovery of two dormant BHs in BH-star binary systems in the galactic field \citep{chakrabarti2022, elbadry2023a, elbadry2023b}.  Combined with follow-up radial velocity measurements to fully characterize the orbit \citep{elbadry2023a, elbadry2023b}, these systems offer new insights into the formation, evolution, and characteristics of BHs in our galaxy.  The first system, Gaia BH1, consists of a Sun-like main sequence (MS) star orbiting a black hole with mass $M_{\mathrm{BH}}=9.62\,\mathrm{M_{\odot}}$, eccentricity $e=0.45$ and orbital period $P=185.6\,$days \citep{elbadry2023b}. Gaia BH2 is a $\sim 1\,\mathrm{M_{\odot}}$ red giant orbiting a a black hole with mass $M_{\mathrm{BH}}=8.9\,\mathrm{M_{\odot}}$, eccentricity $e=0.52$ and orbital period $P=1277\,$days \citep{elbadry2023a}.
According to \cite{elbadry2023b} and \cite{elbadry2023a}, the formation histories of both Gaia BH1 and Gaia BH2 are extremely difficult to explain by isolated binary (IB) evolution, due to the incompatibility between their orbital properties and the outcomes predicted for common envelope evolution. Star clusters (SCs), on the other hand, are highly favored environments for the formation of such systems; not only can dynamical interactions significantly alter the orbital properties of primordial binaries allowing them to follow other evolutionary pathways, but BH-star binaries and their progenitors may also dynamically assemble through binary exchanges and/or multi-body encounters. Despite this, the thin-disk-like Galactic orbits and near-solar metallicities of the Gaia BHs suggest that they most likely did not form dynamically in an old, metal-poor GC.
Young/open SCs, on the other hand, are dynamical systems with Galactic orbits and metallicities perfectly compatible with those of Gaia BHs. Additionally, they are the birthplace of the vast majority of massive stars, which are the progenitors of compact objects \citep[e.g.][] {ladalada2003,portegieszwart2010}. Hence, the majority of BHs in the MW have probably spent the first part of their life in young/open SCs undergoing dynamical interactions.

In this paper, we investigate the dynamical formation of binary systems consisting of a star and a BH in young/open SCs in the MW and via IB evolution. We use a large set of $3\times10^3$ $N$-body simulations of young/open SCs \citep[from][]{dicarlo2020b}, along with population synthesis simulations of IB evolution. We describe the population of BH-star binaries in the MW, predicting the number and properties of the intrinsic population, the expected detections in Gaia DR3, and in future Gaia data releases. Finally, we compare our systems with Gaia BH1 and Gaia BH2 to understand their possible formation pathway, and conclude that both systems (and especially BH2) are more compatible with a dynamical formation scenario.  Throughout this paper, we use ``star'' to refer to every luminous stellar type except for white dwarfs and neutron stars (further discriminating between MS stars and giants when appropriate).

\section{Methods} \label{sec:style}

\subsection{Population of Star Clusters}
\label{sec:pop_sc}

The star cluster simulations used in this paper were performed with the same code and methodology described in \cite{dicarlo2019} and \cite{dicarlo2020b}. Dynamics is treated by the direct summation $N$-Body code \textsc{nbody6++gpu} \citep{wang2015}, coupled with the population synthesis code \textsc{mobse} \citep{mapelli2017,giacobbo2018}, an upgraded version of \textsc{BSE} \citep{hurley2002}. {\sc mobse} includes prescriptions for core-collapse supernovae \citep{fryer2012}, winds of massive stars \citep{giacobbo2018b}, electron-capture supernovae \citep{giacobbo2018c}, natal kicks with fallback \citep{giacobbo2020} and (pulsational) pair instability supernovae \citep{mapelli2020b}. Orbital evolution due to gravitational-wave emission is treated with the equations from \cite{peters1964}. 

In this work, we analyze $3\times 10^3$ simulations of young/open SCs, which are extensively discussed in \cite{dicarlo2020,dicarlo2020b,dicarlo2021}.The initial distribution of stars in space is modeled with fractal initial conditions \citep{kuepper2011, ballone2020, ballone2021, torniamenti2021}, to mimic the asymmetry and clumpiness of star forming regions \citep{goodwin2004}.
The initial stellar mass $M_{\rm SC}$ of every SC ranges from $10^3$ \msun{} to $3\times{}10^4$ \msun{}. Each cluster mass is drawn from a $dN/dM_{\rm SC}\propto M_{\rm SC}^{-2}$ distribution, i.e.~the initial mass function of YSCs in the MW described in \cite{Lada2003}. We calculate the initial half-mass radius $r_{\rm h}$ using the relationship from \citet{markskroupa12}.

The SC initial conditions are generated using {\sc mcluster} \citep{kuepper2011}. We adopt the initial mass function from \cite{Kroupa2001}, with a minimum stellar mass of 0.1 \msun{}  and a maximum stellar mass of 150 \msun{}. The initial total binary fraction is $f_{\mathrm{bin}}=0.4$. Mass ratios are drawn from a distribution $\mathcal{P}(q)\propto{}q^{-0.1}$ (where $q=m_2/m_1\in{}[0.1,1]$, \citealt{sana2012}). All the stars with $\,m\,\ge{}5$~M$_\odot$ are in binary systems, while the stars with $m\,<\,5$~M$_\odot$ are paired stochastically until the initial binary fraction is reached. This is consistent with the multiplicity properties of O/B-type stars (\citealt{sana2012,moe2017}). Eccentricities and orbital periods are drawn from the distributions from \cite{sana2012}.

We set the efficiency of common envelope ejection to $\alpha=5$. We adopt the rapid core-collapse supernovae model, described in \cite{fryer2012}. Natal kicks are randomly drawn from a Maxwellian velocity distribution. A one-dimensional root mean square velocity $\sigma=15\, \mathrm{km/s}$ is adopted for both core-collapse supernovae and electron-capture supernovae.  
We have simulated SCs with three different metallicities: $Z=0.02,$ 0.002 and 0.0002. Each SC is simulated for 100 Myr in a static solar neighbourhood-like tidal field \citep{wang2016}.
From each simulation, we extract all the binaries composed of a BH and a luminous companion which are still in the SC at the end of the simulations, as well as all the binaries of the same type which escape the SC.

\subsection{Synthetic MW of Binaries and Clusters}
\label{sec:synthetic_MW}

Following \cite{chawla2022}, we seed both IBs and SCs following the star-formation history of a cosmological zoom-in simulation of a MW-mass galaxy \citep[specifically the \texttt{m12i} galaxy from the \textit{Latte} simulations,][]{Wetzel2016} from the Feedback In Realistic Environments (\textit{FIRE-2}) suite of galaxies \citep{Hopkins2015,Hopkins2018}.  We extract from the publicly-available data every star particle formed in the simulation (which have masses at formation of $7070M_{\odot}$), including the particle's metallicity and formation time.  These are then binned into low ($[\rm{Fe}/\rm{H}] < -1.5$), intermediate ($-1.5 < [\rm{Fe}/\rm{H}] < -0.5$) and high ($-0.5 < [\rm{Fe}/\rm{H}]$) metallicity bins, designed to match the metallicities of the original \texttt{nbody6++gpu} cluster simulations.  Each metallicity bin is then further divided into 100 equal intervals of cosmic time between 0 and 13.78 Gyr, providing us with the metallicity-dependent star formation rate of our MW-analogue galaxy.

To convert this star formation per unit time into stars and clusters, we begin by assuming that all star formation occurs in clusters \citep{Lada2003}. Since the lowest-mass clusters from \cite{dicarlo2021} begin at $1000M_{\odot}$, we assume that all star formation clumps between $100-1000M_{\odot}$ rapidly dissolve into isolated stars and binaries, while clusters above $1000M_{\odot}$ remain bound and undergo dynamical encounters.\footnote{For an extensive analysis on BH-star binaries in SCs with masses smaller than 1000 \msun{}, we refer to \cite{rastello2023}.} Along with our cluster initial mass function ($dN/dM_{\rm SC}\propto M_{\rm SC}^{-2}$), this suggests that $10\%$ of star formation should occur in open clusters.

We seed our clusters in the \texttt{m12i} galaxy in the following way: for each bin of star formation, we draw randomly with replacement from the cluster sample, subtracting that cluster's mass from the total mass of star particles formed in that bin, until we have completely turned that 10\% of the star formation into clusters.  Each time we draw a cluster, we randomly pick an \texttt{m12i} star particle from the same time/metallicity bin, and assign that particle's exact formation time and present-day position in the \texttt{m12i} galaxy to the BH-star binaries produced by that cluster.  Note that this procedure generates more clusters that present in the original catalog, meaning that many of our dynamical sources are sampled repeatedly. 

The remaining 90\% of star formation is turned into massive binaries in a similar fashion: for each time/metallicity bin, the remaining star formation is converted directly into star particles, assuming the primary masses (that will become BHs) follow a $1/M^{2.3}$ initial mass function (IMF) starting at $18M_{\odot}$ \citep{Kroupa2001} and up to $150\,$\msun{}, while the secondaries are drawn from a uniform mass ratio distribution \citep{Mazeh1992,Goldberg1994} from $0.1M_{\odot}$ up to the primary masses. The binary orbital periods are drawn randomly from a flat-in-log distribution \citep{Abt1983}, while the eccentricities are drawn from a thermal distribution.   Note that these initial conditions (other than the IMF) are different from those assumed in \S \ref{sec:pop_sc}: they are instead chosen to be consistent with \citet{chawla2022}, with a specific focus on forming BH-star binaries.  These systems are then evolved to the present day as described in the next section.  For the \texttt{m12i} galaxy, this procedure yields approximately $9.4\times10^7$ binaries with at least one BH progenitor.

\subsection{Evolving Binaries to the Present Day}
\label{evolvebins}

For consistency with \cite{elbadry2023a,elbadry2023b}, we evolve both IB and SC binaries to the present day using the binary population synthesis code \texttt{COSMIC} \citep{Breivik2020}. The binaries are evolved using the same stellar and binary evolution parameters and recipes used in \citet{dicarlo2020b}, described in \S \ref{sec:pop_sc}, to ensure that binary evolution is treated identically in both populations.
IBs are evolved starting from the Zero-Age Main Sequence (ZAMS) with absolute metallicities of 0.02, 0.002 and 0.0002 (to match the original cluster simulations), from the time when they were formed ($T_{\rm form}$) in the galaxy until the current age of the Universe ($13.78\,\rm{Gyr}$).

The evolution of SC binaries is restarted from the last recorded state of the binary in the $N$-body simulation---that is, either the time of binary ejection or the end of the cluster simulation at 100\,Myr.  
Restarting the evolution of cluster binaries yields additional complications, as the binary evolution in \texttt{COSMIC} for systems having undergone mass loss or accretion depend on the current mass of the star, its effective initial mass, and the change in its effective age due to the mass loss/transfer \citep[referred to as $M(t)$, $M_0$, and the stellar ``epoch'' in][ respectively]{hurley2000}.  $M_0$ and epoch are not part of the default \texttt{nbody6++gpu} output, but since we are primarily interested in BHs and MS stars, this did not present a major difficulty: 
in \texttt{COSMIC}, $M_0$ is equal to $M_t$ for both BHs and MS stars, while the epoch is only relevant for the evolution calculations of MS stars. We assume $\rm{epoch}=0\,\rm{Myr}$ for MS stars when restarting their evolution. 
As a test, we ran the entire population with epoch values of 0 and 100\,Myr (the minimum and maximum values it could have given the SC integration time) and found the difference in the population-level statistics to be negligible (only 1\% systems show a relative difference greater than $\sim$1\% in their final masses and orbital periods between $\rm{epoch}=0\,\rm{Myr}$ and $\rm{epoch}=100\,\rm{Myr}$).  While restarting post-MS stars in \texttt{COSMIC} is substantially more complicated, none of the luminous companion stars evolved beyond the MS before the end of the SC simulations (as opposed to evolving onto the giant branch later in the galactic field).  In our analysis, we distinguish between BH-MS binaries and BH-Giant binaries. We follow the criterion from \citet{Drout:2012} and classify post-MS stars with $T_{\rm eff}\le4800\,K$ as giants. This includes stars with BSE stellar types associated with red giants, core helium burning stars and asymptotic giant branch stars (types 3, 4 and 5).

For our analysis, we only select systems with an age $t_{\mathrm{age}}\geq 100$ Myr. This choice is justified for three reasons: first, our SCs are evolved for 100 Myr, and therefore we are unable to accurately track systems within SCs which formed in the past 100 Myr.  Second, the star-formation history of the \textit{FIRE-2} simulations can be larger than that observed in MW-mass galaxies at late times by a factor of a few \citep[\S 4.5 of][]{Hafen2022}.  This is true in particular for the \texttt{m12i} galaxy we consider, whose $z=0$ star-formation rate \citep[$6 M_{\odot}/\rm{yr}$,][]{Wetzel2016} is $\gtrsim3-4$ times the best observational estimate of star formation in the MW \cite[$1.65\pm0.19M_{\odot}/\rm{yr}$,][]{Licquia2015}. Finally, one of the main focuses of this paper is to understand the formation channels of Gaia BH1 and Gaia BH2, which are both older than $\sim1$ Gyr \citep[and likely significantly older, especially in the case of Gaia BH2,][\S 3.8]{elbadry2023b}.  We briefly discuss the implications of this cutoff (and what our results would look like including such systems) in \S \ref{app1}.

\subsection{Gaia Detectability}\label{sec:gaiadet}

Following \citet{chawla2022}, we further determine whether Gaia can astrometrically resolve each orbit based on the motion of the luminous companion based on an optimistic and pessimistic criteria. In the optimistic case, we assume that Gaia can resolve any orbit for which the star's motion is at least as large as Gaia's single pointing precision, $\sigma_{G}$, when it is projected on the 2d plane of the sky. In the pessimistic case, we require the projected star's orbit to be at least three times as large as $\sigma_{G}$.

\begin{figure}
\begin{center}
\includegraphics[width=1\linewidth]{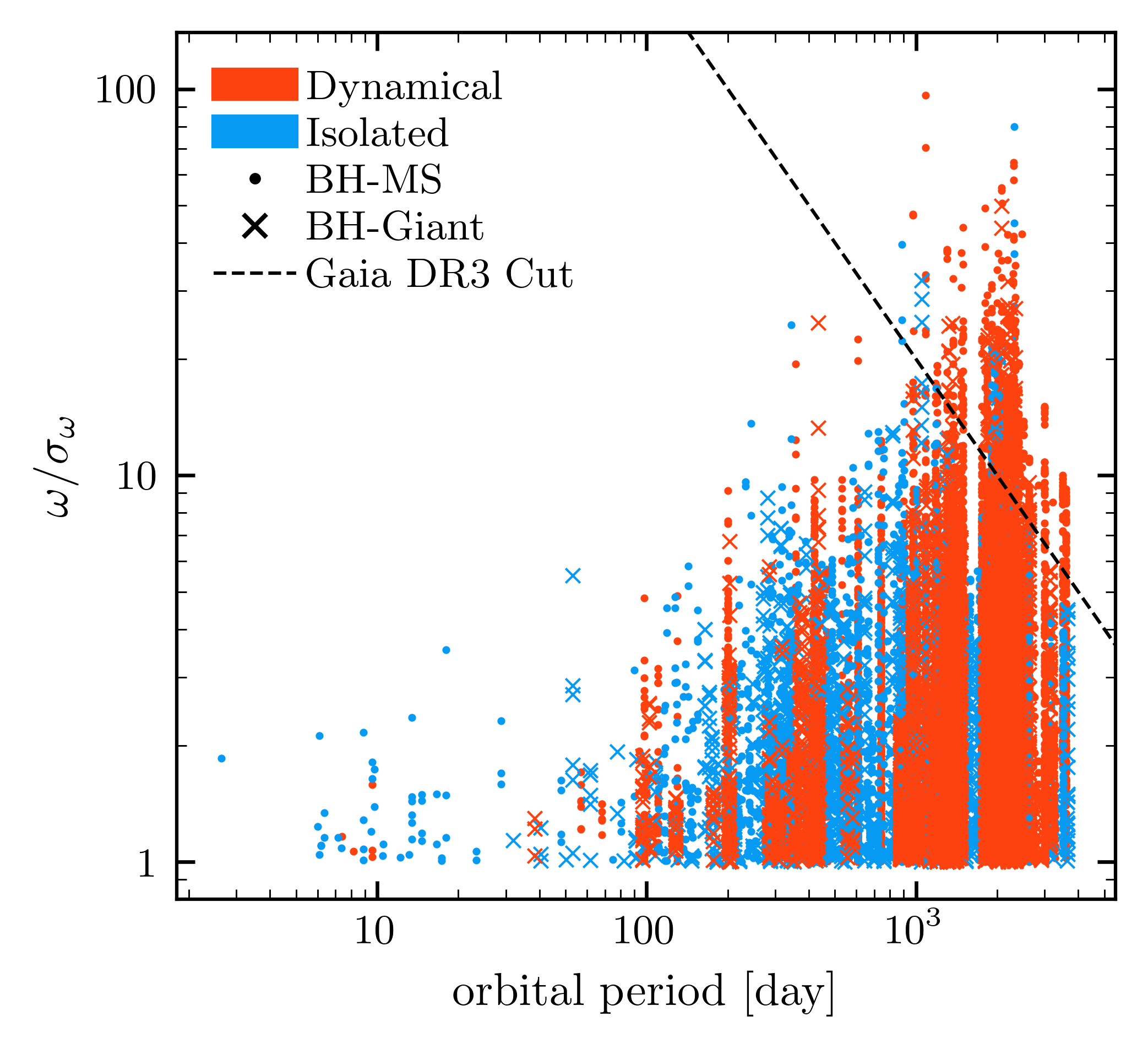}
\caption{\footnotesize The orbital period vs parallax SNR, $\omega/\sigma_\omega$ are shown for BH-star binaries formed in clusters (red) and in isolated binaries (blue) from all of our $100$ MW-like simulation realizations. Dots show BH-MS binaries and crosses show BH-Giant binaries. The dashed black like shows the parallax SNR cut placed as a function of orbital period for Gaia DR3 such that only binaries above the line are included in the data set.  \label{fig:detections}}
\end{center}
\end{figure}

We also consider the cut in the parallax signal-to-noise ratio (SNR), $\omega/\sigma_\omega$, which was applied in in Gaia DR3. This cut was applied as a function of orbital period as shown in Figure~\ref{fig:detections}. The results are shown for the SC BH-star binaries in red and IB binaries in blue.  Dynamically-formed BH-star binaries are preferentially located above the SNR cut while IB BH-star binaries which lie above the cut are mostly due to random-chance placements that are unrealistically close to Earth.

To increase our statistical sample of ``detected'' systems, we consider different positions for Earth in the \texttt{m12i} simulation and combine the results.  We use 100 different equally-spaced starting positions along a ring in the disk with radius $8.5\,\rm{Kpc}$.  For what follows, the results for our synthetic Gaia population are divided by 100 when quoting the number of detectable binaries and all other relevant quantities.

\section{Results}
\subsection{Intrinisic Population}

We first explore the intrinsic population of BH-star binary systems in the galaxy; that is, the full population before considering our Gaia detectability criterion. In all our BH-star systems, the luminous member is either a MS or giant star. The exact numbers of systems per formation channel and stellar type is reported in Table~\ref{tab:table2}. Based on the \texttt{m12i} simulation, we predict that the MW hosts a total of $\sim 2\times 10^5$ BH-star systems from the IB and the young/open SC channels, with $\sim81$\% of the systems formed dynamically and $\sim19$\% formed in isolation. This means that according to our models, SCs produce nearly 4 times more systems than IBs. However, we must emphasize that we have assumed that only 10\% of the star formation occurs in young SCs.  If we take this into account, it is obvious that SCs are \emph{dramatically} more efficient than IBs at producing BH-star systems, with SCs producing $\sim 40$ times as many systems overall, per unit mass, as IB evolution.  This is driven largely by the production of BH-MS star systems; for evolved systems however, the situation is reversed, with IBs producing 2.4 times more BH-Giant binaries than SCs.

\begin{figure}
\begin{center}
\includegraphics[width=0.9\linewidth]{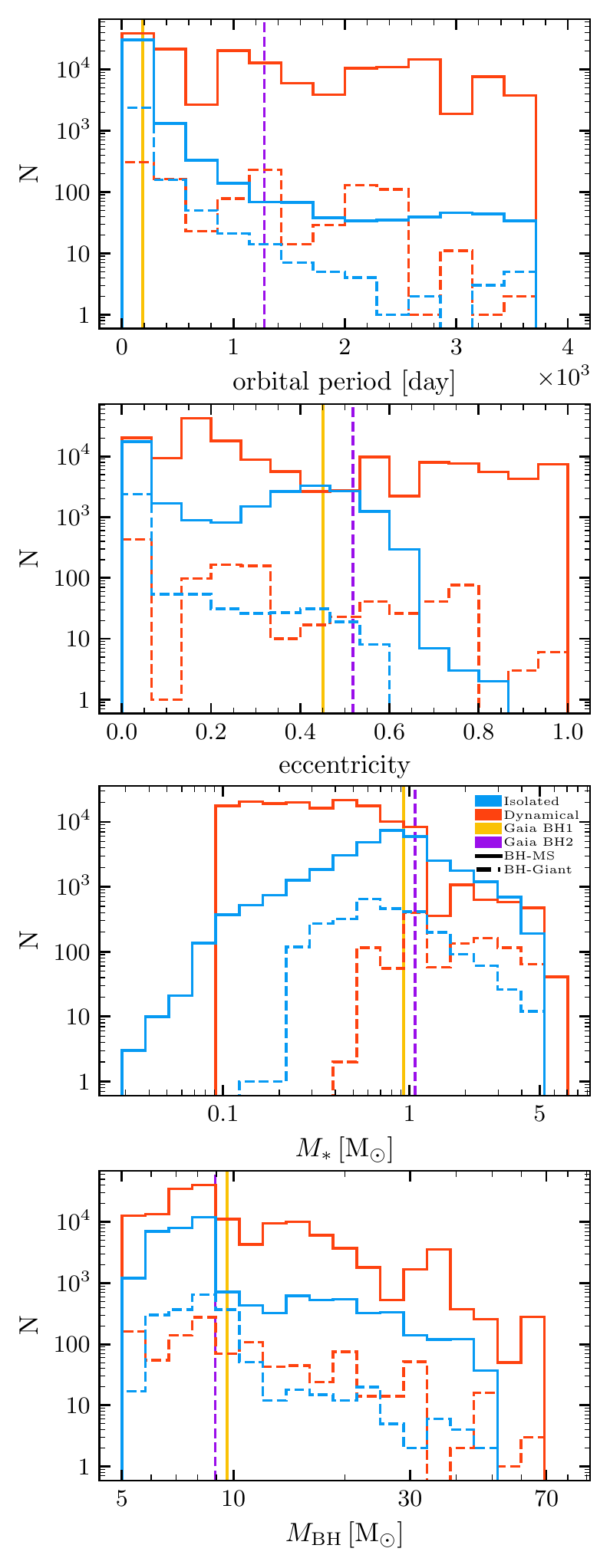}
\caption{\footnotesize \label{fig:hist_intr} Orbital parameters of systems in the intrinsic population. From top to bottom we show the distributions of binary orbital periods, eccentricities, stellar masses $M_*$, and BH masses $M_{\mathrm{BH}}$, with SC binaries shown in red lines and IB shown in blue lines. Solid lines represent BH-MS binaries, while dashed lines show BH-Giant binaries. The yellow and purple vertical lines represent the values of Gaia BH1 and Gaia BH2 respectively.}
\end{center}
\end{figure}

Figure~\ref{fig:hist_intr} shows the distribution of the orbital parameters of the systems in the intrinsic population. SCs are much more efficient at producing systems with large orbital periods for both BH-MS and BH-Giant binaries. IB systems have eccentricities up to $\sim0.84$, while the most eccentric dynamical system has an eccentricity of $\sim0.99$. IBs produce fewer systems with more massive stars: the maximum value of $M_*$ for IB is $\sim5$ \msun{}, while it is $\sim7$ \msun{} for SCs (though this does depend on the 100 Myr cutoff we have employed, as described in \S \ref{app1}). The maximum BH mass is $\sim50$ \msun{} for IBs and $\sim67$ \msun{} for the SC channel. The SC channel is thus able to form systems with eccentricities and masses not accessible from IB evolution.


\begin{table}
\begin{center}

\begin{tabular}{lccc}
\hline
 Channel & All & BH-MS & BH-Giant\\
\hline
\hline
\multicolumn{4}{c}{Formation Efficiency $\mathrm{[M_{\odot}^{-1}]}$} \\
 \hline
 Isolated & $4.55\times10^{-7}$ & $4.21\times10^{-7}$ & $3.40\times10^{-8}$\\
 Dynamical & $1.81\times10^{-5}$ & $1.80\times10^{-5}$ & $1.28\times10^{-7}$\\
 Total & $2.22\times10^{-6}$ & $2.18\times10^{-6}$ & $4.34\times10^{-8}$ \\
\hline
\multicolumn{4}{c}{Intrinsic Population} \\
 \hline
 Isolated & 35229 & 32596 & 2633\\
 Dynamical & 155733 & 154633 & 1100\\
 Total & 190962 & 187229 & 3733\\
\hline
\multicolumn{4}{c}{Gaia DR3 Detections} \\
 \hline
 Isolated & 0.7 & 0.4 & 0.3\\
 Dynamical & 6.6 & 5.2 & 1.4\\
 Total & 7.3 & 5.6 & 1.7\\
\hline
\multicolumn{4}{c}{Gaia Detections (including future data releases)} \\
 \hline
 Isolated & 8-32 & 5-19 & 3-13\\
Dynamical & 53-178 & 40-129 & 13-49\\
 Total & 61-210 & 45-148 & 16-62\\
\hline
\end{tabular}
\caption{\label{tab:table2}Formation efficiency (i.e.~number of systems in the intrinsic population produced per unit simulated stellar mass), number of total BH-star binary systems (intrinsic population), number of expected Gaia DR3 detections and number of pessimistic-optimistic expected Gaia detections including future data releases. Column~1: formation channel of the binary; column~2: value for all BH-star binaries; column~3: value for BH-main sequence binaries only; column~4: value for BH-Giant binaries only. Decimal values come from our average over 100 different observations (see \S \ref{sec:gaiadet}). For Gaia DR3 detections, our optimistic and pessimistic values are identical.}
\end{center}
\end{table}

\subsection{Population Detectable by Gaia}
The Gaia mission provides a unique opportunity to detect a subset of the intrinsic population of BH-star binary systems in the Milky Way.
Table~\ref{tab:table2} presents the number of systems in our data-set according to their detectability by Gaia. The table displays two subsets: the first containing the number of systems detectable by Gaia DR3 in our model, while the second includes systems detectable by both Gaia DR3 and future Gaia data releases. It is evident how SCs are significantly more efficient at producing detectable binaries. In particular, we see that all the 7 expected Gaia DR3 detections are expected to come from the dynamical formation channel. If we also consider detections with future Gaia data releases, we predict that Gaia will detect between 8 and 32 binaries from IBs, and between 53 and 178 binaries from the dynamical channel, i.e.~a factor of $\sim6$ more dynamical systems. In general, our models produce more detectable BH-MS binaries than BH-Giant binaries.

\begin{figure*}
\begin{center}
\includegraphics[width=1.01\linewidth]{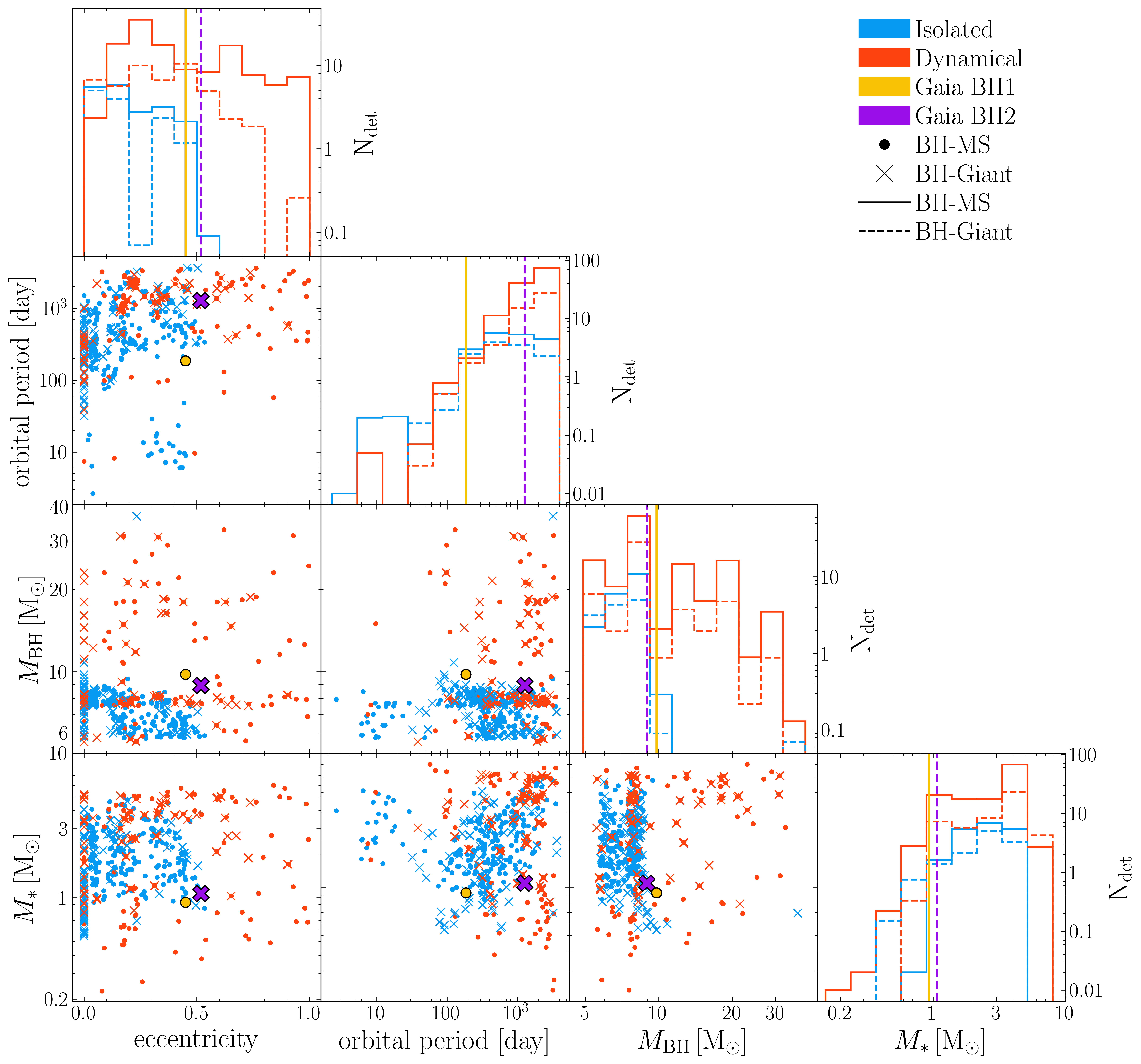}
\caption{\footnotesize \label{fig:cornerdet}  Correlations and distributions of orbital period, eccentricity, mass of the star and mass of the BH of detectable systems. Dynamical systems are shown in red, while isolated systems are shown in blue. Gaia BH1 is shown in yellow, while Gaia BH2 is shown in violet. BH-MS systems are represented by filled circles in the scatter panels and by solid lines in the histogram panels. BH-Giant systems are represented by crosses in the scatter panels and by dashed lines in the histogram panels. Scatter plots show systems from all the 100 observers around the Galaxy (see \S \ref{sec:gaiadet}), while histograms show the number of detections $N_{\mathrm{det}}=N/100$. }
\end{center}
\end{figure*}

Figure~\ref{fig:cornerdet} shows the distributions and correlations of the orbital parameters for the detectable population. It is evident that only SCs are able to produce detectable systems with large eccentricities and with large BH masses. In particular, the detection of a system with $e\gtrsim0.5$ or with $M_{\mathrm{BH}}\gtrsim 10$ \msun{} would be a smoking gun of dynamical formation, as IBs are unable to form detectable systems with these characteristics in our models. The SC channel is more efficient at producing detectable binaries with large orbital periods as well.  However, we reiterate that these results are for systems older than 100 Myr; we explore the implications of this cutoff in \S \ref{app1}.

\subsection{Dynamical Formation and Star Cluster Properties}\label{sec:scprop}
The results from our analysis of the intrinsic and detectable populations strongly suggest that the majority of these systems are formed dynamically in SCs. 
Given that, it is informative to understand how many detectable systems produced by SC evolution come from primordial binaries (i.e.~binaries present in the SC since the beginning of the simulation that may have been altered through dynamical processes) or from binaries that were assembled later dynamically. We find that approximately $94\%$ of detections stem from these dynamically-assembled binaries, while the remaining $6\%$ arise from primordial binaries; $\sim20$\% of the dynamically assembled binaries initially assemble as a star-star system (that later evolves to a BH-star system), while the remaining $\sim80$\% form dynamically from a BH and a star.

\begin{figure}
\begin{center}
\includegraphics[width=0.99\linewidth]{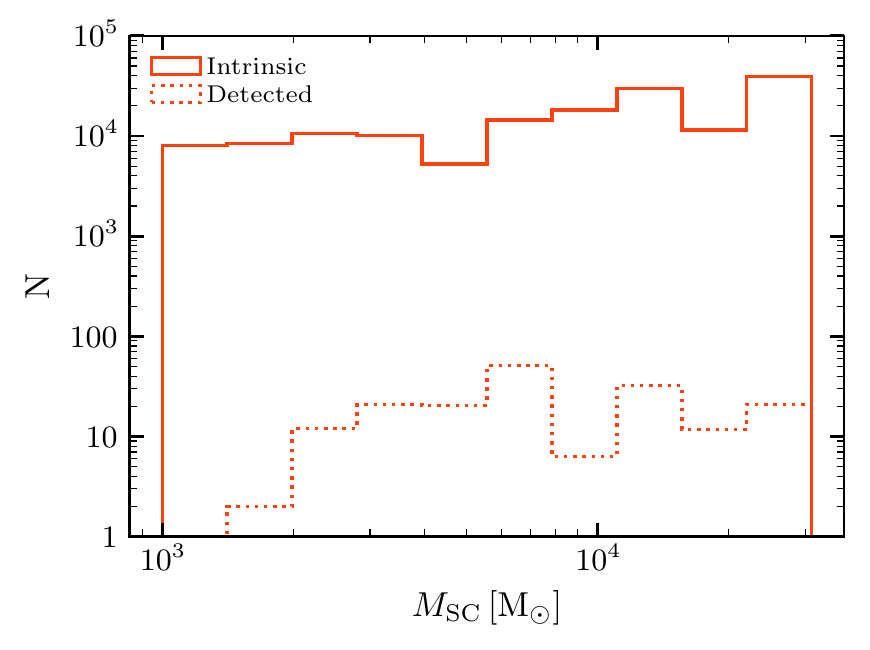}
\caption{\footnotesize \label{fig:msc} Distributions of the initial SC masses $M_{\mathrm{SC}}$ which produce BH-star binaries. The intrinsic population is shown by the solid red line, while the detectable population is shown by the dotted red line.
The initial masses of the simulated SCs are drawn from a $dN/dM_{\rm SC}\propto M_{\rm SC}^{-2}$ distribution \citep{Lada2003}, meaning that we simulated more low mass SCs and less high mass SCs.}
\end{center}
\end{figure}

We can also ask which SCs produce the majority of the BH-star binaries.  In Figure \ref{fig:msc}, we show the number of BH-star binaries produced by SCs of a given mass.  Even though the cluster catalog is sampled following a $\propto M_{\rm SC}^{-2}$ cluster mass function (containing many more low-mass clusters), the intrinsic population exhibits a relatively uniform distribution across SC masses, while the detectable population predominantly arises from SCs with $M_{\mathrm{SC}} \gtrsim 3 \times 10^3$ \msun{}.  For insights into the formation of BH-MS binaries in SCs with lower masses ($M_{\mathrm{SC}}=300-1000$ \msun{}), \cite{rastello2023} show a comprehensive analysis that compares results from low- and high-mass SCs.

The overwhelming majority of observable binary systems originate from SCs with high metallicity (Z=0.02), with $\sim99.84\%$ of them forming in such clusters. This is because of the metallicity-dependent star formation history of the MW.
Massive metal-rich SCs are thus very efficient at producing detectable binaries, and we expect the majority of BH-star systems in the MW and current/future Gaia detections to have formed in such environments.
In a recent study, \cite{tanikawa2023} estimated that the MW hosts around $1.6\times 10^4$ BH-star systems by simulating SCs with $M_{\mathrm{SC}}=1000$ \msun{}. In contrast, our analysis suggests a much larger number of $\sim1.9\times 10^5$ BH-star systems, highlighting the crucial role of more massive SCs to comprehend the formation and evolution of these systems.

We also highlight that $\sim85\%$ of the BH-star binaries formed in SCs are retained by their host SC at the end of the simulations. Despite the relatively low initial mass of the SCs considered in this study, most of them have not undergone complete disruption by the end of the simulations and would likely survive longer \citep[see e.g.][]{torniamenti2022}.  While it is possible that some of the retained binaries may eventually disrupt due to dynamical encounters, it is also highly likely that new BH-star binaries will form as a result of ongoing dynamical interactions.

\section{Forming Gaia BH1 and Gaia BH2}\label{sec:gaiabhs}
\begin{figure*}
\begin{center}
\includegraphics[width=0.75\linewidth]{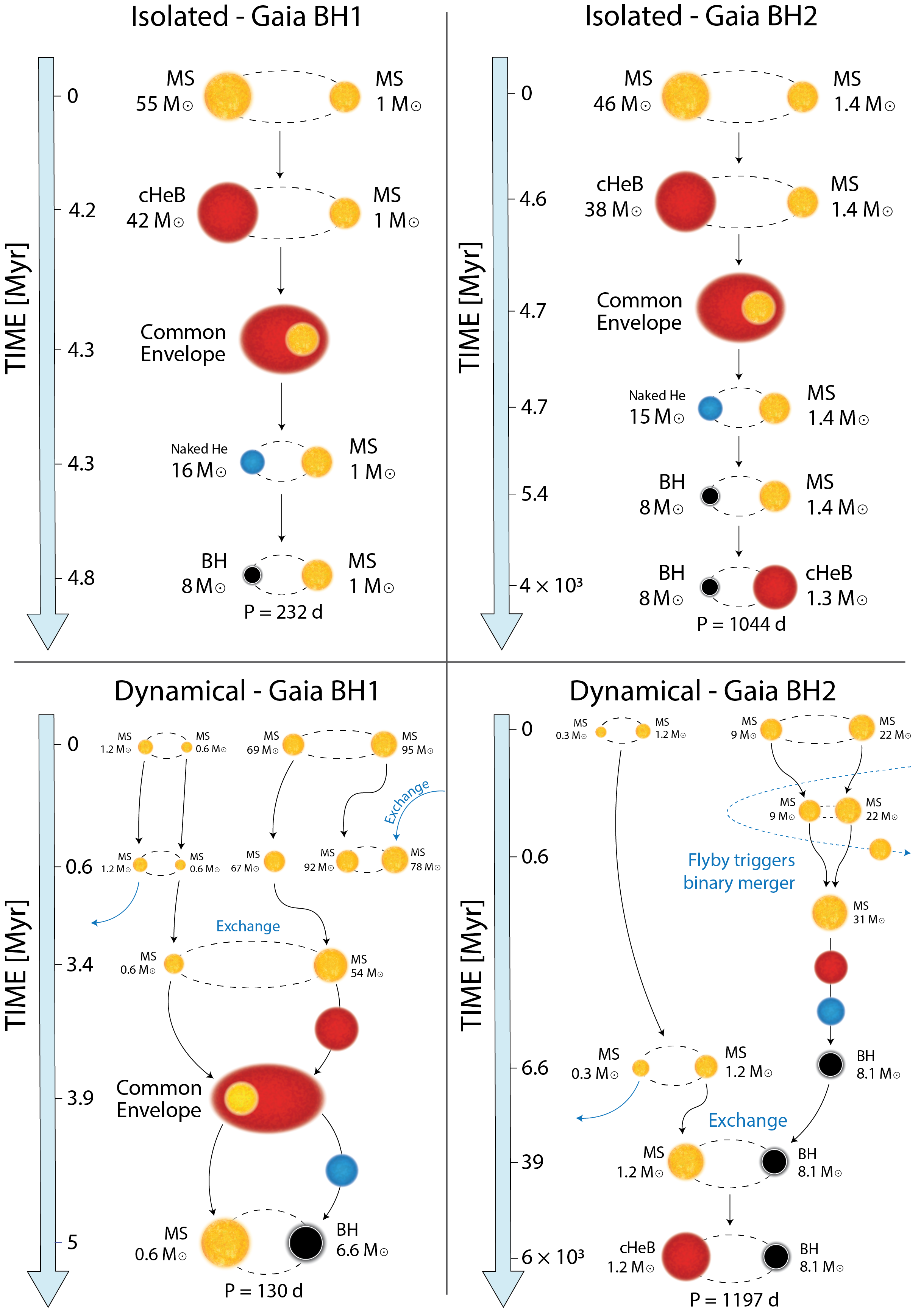}
\caption{\footnotesize \label{fig:formhist}Formation history of the most similar systems to Gaia BH1 (left panels) and Gaia BH2 (right panels) in our simulations. Top panels show systems from the isolated binary evolution channel; bottom panels show systems from the star clusters channel. Main sequence stars (with label MS) are represented as yellow stars; core helium burning stars (label cHeB) are visualized as red stars; naked helium stars (label Naked He) are represented as blue stars; black holes (label BH) are shown as black circles. The mass of each object is shown next to them. The orbital period in days of the final binaries is shown underneath. The time axis and the size of the objects and orbits are not to scale. Even though Gaia BH-like systems can be formed by both the formation channels, our results on rates and ages presented in \S \ref{sec:gaiafc} strongly suggest that the dynamical scenario is the primary formation pathway for the Gaia BHs.}
\end{center}
\end{figure*}

In order to understand how the BH-star binary systems detected by Gaia might have formed, we compare their properties to those of our simulated populations. For both the IB and the SC channel, we find the most similar system to Gaia BH1 and Gaia BH2 in the detectable population, and describe their formation history. The identification of the most similar system entails assessing the fractional differences in key parameters, namely the orbital period, primary mass, and secondary mass, between the simulated binaries and the Gaia BHs. The fractional difference is defined as $|x - x_{\mathrm{GAIA}}| / x_{\mathrm{GAIA}}$, where $x$ is the parameter of the object in the simulated population and $x_{\mathrm{GAIA}}$ is the corresponding parameter of the Gaia BH. The system that exhibits the smallest maximum fractional difference across these parameters (i.e. the one that deviates the least across all the parameters considered) is regarded as the most similar to the Gaia BH. Due to the relatively limited statistical data available for the SC channel, we make the assumption that eccentricities are subject to randomization through dynamical encounters in SCs, where the eccentricity distribution is expected to follow a thermal profile. Thus, we exclude the eccentricity in the evaluation of the most similar systems.
The formation histories of these systems is summarized in Figure~\ref{fig:formhist}.

\subsection{Gaia BH1}
The parameters of the most similar IB system to Gaia BH1 fall within approximately $25\%$ of those of Gaia BH1. The system has an orbital period $P\sim232$ days, $M_{\mathrm{BH}}\simeq 8.4$\msun{} and $M_{*}\simeq 1.05$\msun{}. The initial conditions of the system are $P\simeq3486$ days, $M_{\mathrm{1}}\simeq 55.1$\msun{}, $M_{2}\simeq 1.04$\msun{} and a metallicity of $Z=0.02$. The system undergoes a common envelope episode that significantly shrinks its orbit, and then turns into a BH-MS binary after 4.8 Myr. The binary then evolves unperturbed to the present day. The system has an age of $\sim 1.5$ Gyr.

The closest system from the SC channel has the parameters within $\sim32\%$ of Gaia BH1.
 The system has an orbital period $P\sim130$ days, $M_{\mathrm{BH}}\simeq 6.6$\msun{} and $M_{*}\simeq 0.6$\msun{}. The system assembles dynamically inside a SC with metallicity $Z=0.02$ and initial mass $M_{\mathrm{SC}}=3204$ \msun{}. The binary forms $\sim 5$ Myr after the beginning of the simulation. The primary turns into a giant, triggering a common envelope episode that shrinks the orbit. Afterwards, the envelope is ejected, and the primary turns into a BH. $\sim 10$ Myr after its formation, the system is dynamically ejected from the SC. The age of this system is $\sim3$ Gyr.

Both the IB and SC models struggle to reproduce a system that very closely resembles Gaia BH1. \cite{rastello2023} suggests that the formation of Gaia BH1 might be explained by less massive SCs, with masses ranging from 300\msun{} to 1000\msun{}. These lower mass star clusters could provide a favorable environment for the formation of a system with properties similar to Gaia BH1.
 Based on our analysis of the age distribution and of our detection rates for BH-MS systems in Gaia DR3, we draw the conclusion that the dynamical formation channel is the most favorable for Gaia BH1, as we better explain in \S \ref{sec:gaiafc}. 
\subsection{Gaia BH2}
The closest IB system to Gaia BH2 is a BH-Giant binary with $P\sim1044$ days, $M_{\mathrm{BH}}\simeq 8.0$\msun{} and $M_{*}\simeq 1.3$\msun{}. The values lie within $\sim19\%$ of the parameters of Gaia BH2. The initial conditions of the system are $P\simeq3967$ days, $M_{\mathrm{1}}\simeq 45.8$\msun{}, $M_{2}\simeq 1.36$\msun{}. The system has a metallicity of $Z=0.02$, and undergoes a common envelope episode throughout its evolution. The system has an age of $\sim 4.1$ Gyr, which is not compatible with the age of Gaia BH2.

The best candidate from the SC channel has the parameters within $13\%$ of the Gaia BH2 parameters.
 The system has an orbital period $P\sim1197$ days, $M_{\mathrm{BH}}\simeq 8.1$\msun{} and $M_{*}\simeq 1.2$\msun{}. The binary is dynamically assembled in a SC with metallicity $Z=0.02$, initial mass $M_{\mathrm{SC}}=14284$ \msun{} and an age of $\sim6$ Gyr. The system forms as a BH-MS binary $\sim 39$ Myr after the beginning of the simulation, and it later evolves to a BH-Giant. The system is retained by its host SC at the end of the simulation. 

Both channels produce systems that exhibit close parameter matches to Gaia BH2. Among these, the dynamical scenario yields the closest resemblance. If we also take into account that Gaia BH2 is estimated to be older than 5 Gyr, and that our Gaia DR3 detections for BH-Giant systems come entirely from SCs, we conclude that Gaia BH2 has likely formed via the dynamical formation channel. We provide a more extensive analysis on this in \S \ref{sec:gaiafc}. 

\section{Discussion}
\subsection{Dynamical versus Isolated}\label{sec:dyniso}
We see from Table~\ref{tab:table2} that the dynamical channel produces $\sim4.3$ times more BH-star systems than the isolated channel. Since we consider that only 10\% of the total star formation occurs into SCs, we find that the number of systems produced per unit stellar mass by SCs is $\sim40$ times larger than the number of systems produced by isolated binary evolution.
This discrepancy can be attributed to several mechanisms. Stellar dynamics in SCs significantly affects the formation and evolution of binaries. Multi-body encounters can dynamically assemble BH-star or star-star binaries which can later evolve into BH-star binaries, resulting in combinations of orbital parameters that are inaccessible in IB evolution.
Furthermore, three-body encounters between a binary and a single object can modify the eccentricity and orbital period of binaries, triggering or preventing mass transfer episodes (e.g., Roche-lobe overflow and common envelope evolution) that may or may not occur if the binary evolved in isolation. Common envelope episodes significantly reduce the orbital period of binaries. From Figure~\ref{fig:hist_intr} and Figure~\ref{fig:cornerdet}, we see how the dynamical channel is extremely more efficient at producing binaries with larger orbital periods. Indeed, we find that 99.4\% of all the simulated IB systems which become BH-star binaries undergo at least one common envelope episode throughout their evolution. This can be easily avoided in SCs; as we said in \S \ref{sec:scprop}, the majority of detectable SC binaries dynamically assembles as a BH-star binary already, which means that they do not undergo a common envelope episode\footnote{The only way for such binaries to undergo a common envelope episode and still be classified as BH-star binaries is if the following sequence of events occurs: i) the star becomes a giant; ii) its expansion triggers a common envelope; iii) the envelope is ejected by the BH companion, leaving behind a BH and a naked He star. We do not find any BH-naked He star binary in our populations.}.  
The dynamical channel is thus not only more efficient, but can also leave unique fingerprints on the properties of the binaries.

\subsection{Gaia BHs formation channels}\label{sec:gaiafc}
Our models provide several hints on how the dynamical scenario is a more favorable formation pathway for Gaia BH1 and Gaia BH2. 
According to our results in Table \ref{tab:table2}, all the expected Gaia DR3 detections come from SCs, suggesting that Gaia BHs have more likely formed dynamically.
In order to form systems like Gaia BH1 or Gaia BH2 through IB evolution, \cite{elbadry2023b} and \cite{elbadry2023a} highlight the necessity of assuming an exceedingly large and potentially unrealistic value for the common envelope efficiency ($\alpha \approx 12.8$ for Gaia BH1). Lower values of $\alpha$ would have led to the merger of the binary components, preventing the formation of Gaia BHs through this channel. As discussed in \S \ref{sec:dyniso}, dynamics offers an alternative pathway for the formation of BH-star systems that completely bypasses the common envelope phase.

The age distribution of the detected systems is shown in Figure~\ref{fig:ages}. The efficiency of both channels is comparable at 0.1 Gyr, but the dynamical channel is much more efficient at forming detectable systems with older ages. Gaia BH1 has an age $\gtrsim$1 Gyr, while Gaia BH2 has an age $\gtrsim$5 Gyr, indicating that both Gaia BHs, particularly Gaia BH2, have likely formed dynamically, according to our models.

\cite{rastello2023} show that both lower mass (300-1000\msun{}) and higher mass (1000-30000 \msun{}) young SCs are efficient in forming Gaia BH1-like systems.
Gaia BHs may also have formed via other channels not taken into account in this study, such as formation in isolated hierarchical triples or hierarchical triples formed in SCs \citep[see e.g.][]{trani2022}. Another plausible channel could be the dynamical formation in a globular cluster (GC); while the Galactic orbit of both Gaia BHs is not aligned with any GC in the MW \citep{elbadry2023b, elbadry2023a}, these systems might have dynamically formed in GCs which have already completely disrupted.

\subsection{BH-star ages and the 100 Myr age cutoff}\label{app1}

\begin{figure}
\begin{center}
\includegraphics[width=0.99\linewidth]{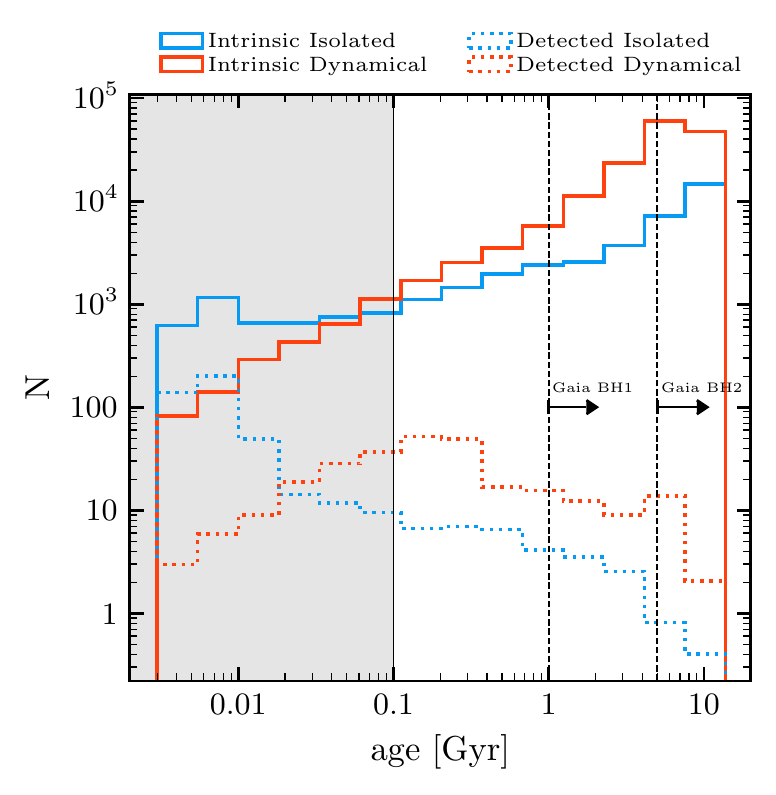}
\caption{\footnotesize \label{fig:ages}  Distributions of the ages of systems in our datasets. The intrinsic population is shown by solid lines, while Gaia detectable systems are shown by dotted lines. Dynamical systems are shown in red, while isolated systems are shown in blue. The two dashed vertical lines show the lower limits of the age of Gaia BH1 and Gaia BH2. The grey shaded area on the left represents the systems not taken into account due to the age cutoff described in \S \ref{evolvebins}.}
\end{center}
\end{figure}  

\begin{table}
\begin{center}
\begin{tabular}{lccc}
\hline
Channel & All & BH-MS & BH-Giant\\
\hline
\hline
\multicolumn{4}{c}{Formation Efficiency $\mathrm{[M_{\odot}^{-1}]}$} \\
 \hline
 Isolated & $5.13\times10^{-7}$ & $4.79\times10^{-7}$ & $3.47\times10^{-8}$\\
 Dynamical & $1.84\times10^{-5}$ & $1.83\times10^{-5}$ & $1.30\times10^{-7}$\\
 Total & $2.30\times10^{-6}$ & $2.26\times10^{-6}$ & $4.42\times10^{-8}$ \\
\hline
\multicolumn{4}{c}{Intrinsic Population} \\
 \hline
 Isolated & 39737 & 37048 & 2689\\
 Dynamical & 158329 & 157214 & 1115\\
 Total & 198066 & 194262 & 3804\\
\hline
\multicolumn{4}{c}{Gaia DR3 Detections} \\
 \hline
 Isolated & 87 & 82 & 5\\
 Dynamical & 14 & 12 & 2\\
 Total & 101 & 94 & 7\\
\hline
\multicolumn{4}{c}{Gaia Detections (including future data releases)} \\
 \hline
 Isolated & 286-455 & 272-430 & 14-25\\
Dynamical & 88-272 & 73-220 & 15-52\\
 Total & 374-727 & 345-650 & 29-77\\
\hline
\end{tabular}
\caption{\label{tab:table3}Same as Table \ref{tab:table2}, but without the 100 Myr formation time cutoff.}
\end{center}
\end{table}

Throughout this paper, we have argued that SCs dominate the production of old BH-star systems in the MW.  We show this explicitly in Figure \ref{fig:ages}, which displays the distribution of ages of the systems, and reveals that the dynamical formation channel appears to be more efficient at producing older systems, particularly those older than 1 Gyr. 

Of course, it is also obvious in Figure \ref{fig:ages} that our 100 Myr cutoff has removed a large number of young systems, particularly from the IB channel, that contribute significantly to the detectable population.  For comparison, we present the results obtained without applying the 100 Myr age cutoff (which was justified in \S \ref{evolvebins}), starting with Table~\ref{tab:table3} which reports the number of systems in the intrinsic and detectable populations. Although there is an increase in the number of systems in the isolated intrinsic population, the dynamical channel remains dominant, constituting approximately $80\%$ of the total intrinsic population.

\begin{figure}
\begin{center}
\includegraphics[width=0.88\linewidth]{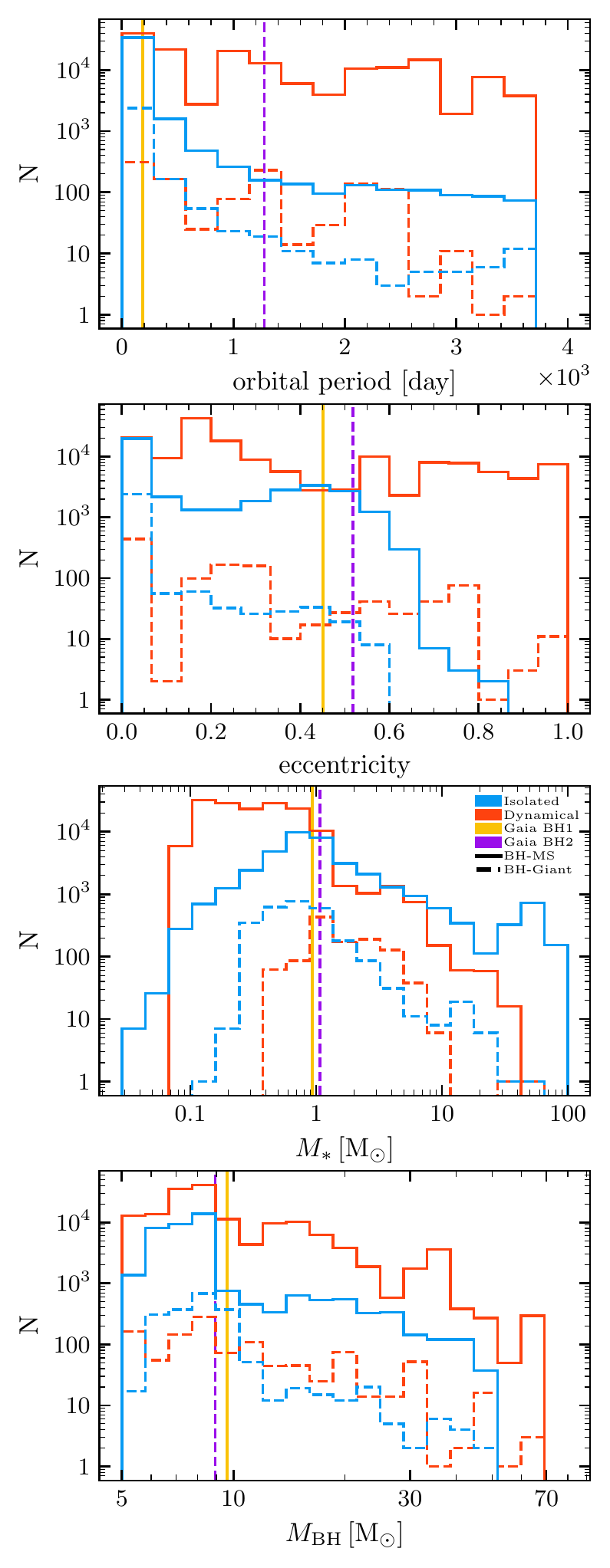}
\caption{\footnotesize \label{fig:nocutoffdist} Distributions of orbital parameters of systems in the intrinsic population, including systems with age $< 100$ Myr. From top to bottom: orbital period, orbital eccentricity $e$, star mass $M_*$, BH mass $M_{\mathrm{BH}}$. SC binaries are shown by red lines, IB are shown by blue lines. Solid lines represent BH-MS binaries, while dashed lines show BH-Giant binaries. The yellow and purple vertical lines represent the values of Gaia BH1 and Gaia BH2 respectively.}
\end{center}
\end{figure}

\begin{figure}
\begin{center}
\includegraphics[width=0.88\linewidth]{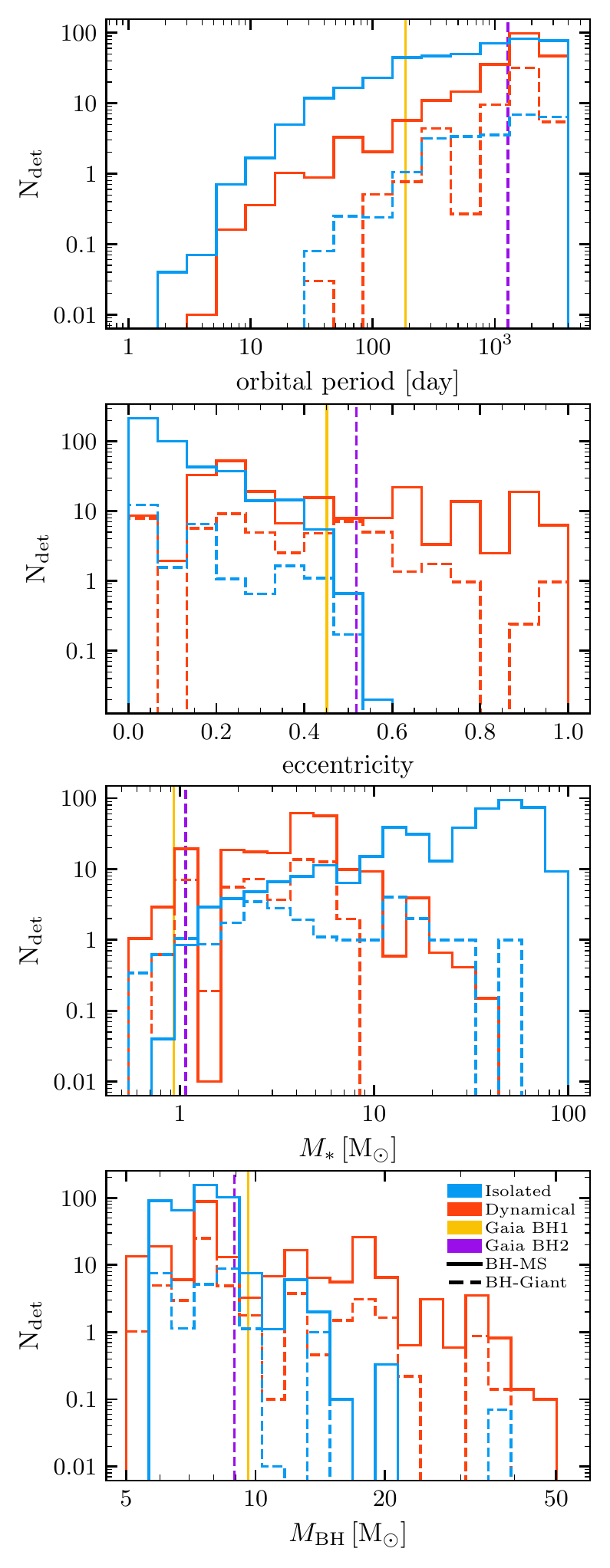}
\caption{\footnotesize \label{fig:nocutoffdistdet} Same as Figure \ref{fig:nocutoffdist}, but for systems detectable in future Gaia data releases.}
\end{center}
\end{figure}

However, we observe a significant increase in the number of isolated detectable systems. The expected number of IB detections by Gaia DR3 increases from 0 to 87, while the total number of IB Gaia detections increases from 8-32 to 286-455. Conversely, the number of detectable systems from the dynamical channel barely increases. This disparity can be attributed to the fact that systems in the isolated population are predominantly very young (see Figure~\ref{fig:ages}) and preferentially host bright massive stars that are significantly easier to detect. This trend is evident from Figure~\ref{fig:nocutoffdist} and Figure~\ref{fig:nocutoffdistdet}, which illustrate the significant impact of the cutoff on the $M_*$ distribution. The difference comes almost entirely from a huge population of young massive O-stars. The IB detectable systems younger than 20 Myr have an average BH mass of 8 \msun{} and an average star mass of 41 \msun{}, and constitute $\sim86\%$ of the total number of detectable systems in the whole IB catalogue without the 100 Myr age cutoff.  We emphasize that our predictions for the formation channels of both Gaia BH1 and Gaia BH2 should remain unaffected by this cutoff, as they both are older than 1 Gyr.




\section{Conclusions} \label{sec:conc}
The MW is very likely populated by a large number of stellar BHs, many of which probably reside in binary systems. The recent discovery of two BH-star binaries in Gaia DR3 has sparkled new interest in understanding the formation channels and population characteristics of such systems.
In this paper, we have investigated the formation of BH-star binaries in young SCs and via IB evolution.

According to our simulations, the MW harbors a total of $\sim 2\times 10^5$ BH-star systems of which $\sim 81$\% formed dynamically and $\sim 19$\% formed in isolation. We find that dynamical formation in SCs is nearly 40 times more efficient per unit stellar mass at producing BH-star binaries compared to isolated binary evolution.
Dynamical systems tend to have larger orbital periods and eccentricities than the isolated ones.
We expect that a total of 7 BH-star systems are present in the Gaia DR3 data, all of which come from the dynamical channel.
We also predict between 61 and 210 detections from the whole Gaia mission, $\sim83$\% of which come from SCs.
Overall, dynamics enhances dramatically the number of BH-star binaries, both in the intrinsic and the detectable populations.

We compare our detectable populations with Gaia BHs, and we conclude that our models support the dynamical scenario as the primary formation pathway for Gaia BH1 and Gaia BH2.  Our results suggest that identifying systems with an an eccentricity greater than $\sim0.5$, or with black hole mass $M_{\mathrm{BH}}$ exceeding 10 \msun{} would provide compelling evidence for their dynamical formation.  With new detections from future Gaia data releases expected over the next few years, these results will help disentangle the formation channels of BH-star systems and the characteristics of the BH population in the MW.

Finally, our analysis also revealed the presence of some BH-Giant systems with 3 \msun{} $<M_{\mathrm{BH}}<$ 5 \msun{}, i.e.~below the minimum BH mass allowed by the rapid supernova mechanism \citep{fryer2012}. These peculiar systems form through accretion of matter by neutron stars from their binary companions, leading to the formation of BHs through accretion-induced collapse (AIC). None of these systems are detectable by Gaia according to our analysis, but we have not yet compared these systems (identified only in the IB channel) to their dynamical counterparts.  Efforts to better characterize these systems are currently underway.

\begin{acknowledgments}
We thank Michela Mapelli, Sara Rastello, Giuliano Iorio and all the members of the DEMOBLACK group for the useful discussions. This work was supported by NSF Grant AST-2009916 to The University of North Carolina at Chapel Hill and Carnegie Mellon University. The $N$-body simulations of young star clusters adopted here are an open-data product of the European Research Council (ERC) Consolidator grant DEMOBLACK, contract no. 770017 (PI: Michela Mapelli). CR acknowledges support from a Charles E.~Kaufman Foundation New Investigator Research Grant, an Alfred P.~Sloan Research Fellowship, and a David and Lucile Packard Foundation Fellowship. The Flatiron Institute is supported by the Simons Foundation.
\end{acknowledgments}

%







\bibliography{gaiabhstar}{}
\bibliographystyle{aasjournal}



\end{document}